\documentstyle[epsfig]{aipproc}

\newcommand{\vzrms}{v_{\rm rms,0}}
\newcommand{\vrms}{v_{\rm rms}}
\newcommand{\beq}{\begin{equation}}
\newcommand{\eeq}{\end{equation}}
\def\ga{\;\rlap{\lower 2.5pt
 \hbox{$\sim$}}\raise 1.5pt\hbox{$>$}\;}
\def\la{\;\rlap{\lower 2.5pt
   \hbox{$\sim$}}\raise 1.5pt\hbox{$<$}\;}

\begin{document}
\title{Warm Dark Matter, Small Scale Crisis, and the High Redshift Universe}

\author{Zoltan Haiman$^*$\thanks{Hubble Fellow}, Rennan Barkana$^{\dagger}$
and Jeremiah P. Ostriker$^*$}
\address{$^*$ Princeton University Observatory, Princeton, NJ 08544\\
$^{\dagger}$Canadian Institute for Theoretical Astrophysics, 
60 St. George Street \#1201A, Toronto, Ontario, M5S 3H8, Canada}

%\lefthead{LEFT head}
%\righthead{RIGHT head}
\maketitle

\begin{abstract}
Warm Dark Matter (WDM) models have recently been resurrected to
resolve apparent conflicts of Cold Dark Matter (DM) models with
observations.  Endowing the DM particles with non--negligible
velocities causes free--streaming, which suppresses the primordial
power spectrum on small scales.  The choice of a root-mean-square
velocity dispersion $\vzrms \sim 0.05$ km/s at redshift $z=0$
(corresponding to a particle mass $m_X\sim 1~$ keV if the WDM
particles are fermions decoupling while relativistic) helps alleviate
most, but probably not all, of the small--scale problems faced by CDM.
An important side--effect of the particle velocities is the severe
decrease in the number of collapsed halos at high redshift. This is
caused both by the loss of small--scale power, and by the delay in the
collapse of the smallest individual halos (with masses near the
effective Jeans mass of the DM).  The presence of early halos is
required in order (1) to host either early quasars or galaxies that
can reionize the universe by redshift $z=5.8$, and (2) to allow the
growth of the supermassive black hole believed to power the recently
discovered quasar SDSS 1044-1215 at this redshift.  We quantify these
constraints using a modified Press-Schechter formalism, and find
$\vzrms \la 0.04$ km/s (or $m_X \ga 1~$keV).  If future observations
uncover massive black holes at $z \ga 10$, or reveal that reionization
occurred at $z\ga 10$, this could conclusively rule out WDM models as
the solution to the small--scale crisis of the CDM paradigm.
\end{abstract}

\section*{Introduction}  
\label{sec:Intro}

The currently favored model of hierarchical galaxy formation in a
universe dominated by cold dark matter (CDM) has been very successful
in matching observations of the density distribution on large scales.
However, recently some small-scale shortcomings of this model have
appeared, as summarized elsewhere in these proceedings (see also
\cite{sk00} for a recent review).  Although the observational  
significance of the discrepancies is still disputed, and astrophysical
solutions involving feedback may still be possible, the accumulating
tension with observations has focused attention on solutions involving
the particle properties of dark matter.  Proposals have included
self-interacting dark matter \cite{ss00}, adding a repulsive
interaction to gravity \cite{goodman,peebles}, the quantum--mechanical
wave properties of ultra--light dark matter particles \cite{fuzzy},
and a resurrection of warm dark matter (WDM) models
\cite{colin00,larsen,Bode}.  By design, a common feature of models
that attempt to solve the apparent small-scale problems of CDM is the
reduction of fluctuation power on small scales.  In the CDM paradigm,
structure formation proceeds bottom-up: the smallest objects collapse
first, and they subsequently merge together to form larger objects. It
then follows that the loss of small-scale power modifies structure
formation most severely at the highest redshifts; in particular, the
number of self--gravitating objects at high redshift is reduced.

A strong reduction in the abundance of high-redshift objects could be
in conflict with observations. First, the lack of a Gunn-Peterson
trough in the spectrum of the bright quasar SDSS 1044-1215 at redshift
$z=5.8$ \cite{f00} implies that the hydrogen in the intergalactic
medium (IGM) was highly ionized prior to this redshift.  The most
natural explanation for reionization is photo-ionizing radiation
produced by an early generation of stars or quasars \cite{me01}. The
sources of reionization reside in halos that have masses in the range
corresponding to dwarf galaxies --- the mass scale on which power
needs to be reduced relative to CDM models.  Second, if SDSS 1044-1215
is unlensed and radiating at or below the Eddington limit, its unusual
intrinsic brightness implies that it is powered by an exceptionally
massive black hole (BH).  The growth of this BH, out of a stellar
remnant seed, requires \cite {hl00} that a host halo be present at a
sufficiently high redshift ($z \gg 5.8$).

In this contribution, we briefly review the status of WDM models in
resolving the problems of CDM, and then examine new constraints that
arise on WDM models from the high redshift universe.  We focus on WDM
models, although similar constraints would apply to other
modifications of the CDM paradigm that reduce the small-scale power.
The cosmological parameters we adopt, based on present large scale
structure data \cite{concord}, are
$(\Omega_0,\Omega_\Lambda,\Omega_{\rm
b},h,\sigma_{8},n)=(0.3,0.7,0.045,0.7,0.9,1)$.  The details of the
study described here can be found in \cite{wdmpaper}.

\section*{Warm Dark Matter Models}  
\label{sec:WDM}

The WDM is assumed to be composed of particles of about $\sim 1$ keV
mass (compared to $\sim 1$ GeV in CDM, or $\sim 10$ eV in Hot Dark
Matter models).  The thermal velocities of the particles cause free
streaming out of overdense regions, smoothing out small--scale
fluctuations, leading to a small-scale cutoff in the linear power
spectrum.  In addition, the thermal velocities act similarly to
pressure, and inhibit the growth of low--mass perturbations.  One
example of WDM is fermionic particles that decouple in the early
universe while relativistic and in thermal equilibrium \cite{text}.
To produce a given contribution $\Omega_X$ to the cosmological
critical density, the required particle mass $m_X$ is then determined
by $m_X n_X \propto \Omega_X h^2$, where the present number density
$n_X$ of WDM particles follows from their chosen r.m.s. velocities.
This yields a relation between particle mass and r.m.s. 
velocity dispersion \cite{Bode},
\beq 
\vrms(z)= 0.0437 \left( 1+z \right)
\left( \frac{ \Omega_X h^2} {0.15}
\right)^{1/3} \left( \frac{g_X} {1.5} \right)^{-1/3} \left( \frac{m_X}
{\rm 1~keV} \right)^ {-4/3}~{\rm km~s^{-1}}, 
\eeq 
where $g_X$ is the
effective number of degrees of freedom of WDM. The comoving cutoff
scale $R_c$, where free--streaming reduces the power spectrum to
half of its value in CDM, is given by
\begin{eqnarray} 
R_c & = & 0.201 \left( \frac{ \Omega_X h^2} {0.15}
\right)^{0.15} \left( \frac{g_X} {1.5} \right) ^{-0.29} \left(
\frac{m_X} {\rm 1~keV} \right)^{-1.15}~{\rm Mpc}\ \nonumber \\ & = &
0.226 \left( \frac{ \Omega_X h^2} {0.15} \right)^{-0.14} \left(
\frac{\vzrms} {\rm 0.05~km/s} \right)^{0.86}~{\rm Mpc}\ . 
\label{eq:Rc} 
\end{eqnarray}
The length scale $R_c$ corresponds to a characteristic mass scale $M_c$,
\beq 
M_c = 1.74 \times 10^8
\left( \frac{ \Omega_0 h^2} {0.15} \right) \left( \frac{ R_c} {\rm
0.1~ Mpc} \right)^3~M_{\sun}\ . 
\eeq
The predictions of WDM models can be expected to differ from CDM on
scales below $R_c$ or $M_c$.  In addition, it is useful to define an
``effective Jeans mass'' for WDM, at which the pressure corresponding
to the r.m.s. particle velocities balances gravity,
\begin{eqnarray} 
M_J
& = & 3.06 \times 10^8 \left( \frac{g_X} {1.5} \right)^{-1} \left(
\frac{ \Omega_X h^2} {0.15} \right)^{1/2} \left( \frac{m_X} {\rm
1~keV} \right)^{-4} \left( \frac{1+z_i}{3000} \right)^{3/2}~M_{\sun}\
\nonumber \\ 
& = & 4.58 \times 10^8 \left( \frac{ \Omega_X h^2} {0.15}
\right)^{-1/2} \left( \frac{\vzrms} {\rm 0.05~km/s} \right)^{3} \left(
\frac{1+z_i}{3000} \right)^{3/2}~M_{\sun}\ . \label{MJeans}
\end{eqnarray} 
We have verified these scalings using spherically symmetric,
one--dimensional collapse simulations \cite{wdmpaper}.  The growth of
perturbations on scales below $M_J$ is slowed down by the ``pressure''
of WDM.  Although this effect is irrelevant in most discussions of WDM
models at lower redshifts (due to the smallness of $M_J$), we find
that it is important in the context of reionization.  The inhibited
growth of perturbations in the linear regime results in a delay in the
final virialization epoch of the individual, low--mass halos, that
first condense out in the universe.

\begin{table}
\caption{Scorecard of WDM models.}
\label{table1}
%\begin{tabular}{lrrr}
%\begin{tabular}{lccc}
%\begin{tabular}{lddd}
\begin{tabular}{llll}
   CDM Problem\tablenote{Note that WDM particles with $m_X\la 750$ eV
   would contradict the power spectrum of the Ly$\alpha$
   forest\cite{vijay00}. A similar limit follows from the
   maximum observed phase space density in cores of dwarf
   spheroidals\cite{hogan,dalcanton}.}  &
\multicolumn{1}{c}{WDM mass}   & 
\multicolumn{1}{c}{Status} &   
\multicolumn{1}{c}{References} \\
\tableline
 Milky Way satellites  & 0.6$-$1.5 keV   & $\surd$ & \cite{colin00,Bode,croft} \\
 Halo concentration    & 0.6$-$1.5 keV   & $\surd$ & \cite{colin00,Bode,hogan} \\
 Inner density profile & $<$0.6 keV ?  & {\bf ?}& \cite{colin00,hogan} \\
 No dwarfs in voids    & $<$1.5 keV  & {\bf ?} & \cite{Bode,peeblesvoid}   \\
 Angular momentum      & 0.5$-$0.8 keV & x{\bf ?}& \cite{larsen,kravtsov}    \\
\end{tabular}
\end{table}

\section*{Summary of Current Observations}  

The current status of WDM models in resolving various problems of CDM is
summarized in Table \ref{table1}.  Based on analytical arguments, as well as
numerical simulations, WDM particles with a mass in the range $0.6$ keV $\la
m_X\la$ 1.5 keV can strongly suppress the number of satellite halos of the
Milky Way, and produce low halo concentration parameters for both dwarfs, and
Milky Way-sized galaxies \cite{colin00,Bode}.  It remains unclear whether WDM
can fully resolve some other, related problems.  One of the major successes of
the CDM paradigm is the interpretation of the statistics of the Ly$\alpha$
forest. An analysis of the Ly$\alpha$ forest power spectrum at redshift
$z\approx 3$ finds that WDM models with $m_X<0.75$ keV would spoil this success
\cite{vijay00}.  A similar limit from the maximum observed phase space density
in dwarf spheroidal galaxies. To solve some of the problems with CDM, lower
masses might be needed. In existing simulations for $m_x>$ 0.6 keV, the inner
slope of the density profiles of normal disk galaxies do not appear to flatten
sufficiently to produce the observed, nearly constant cores, although these
simulations do not yet probe the relevant innermost regions
reliably\cite{colin00}.  A similar situation arises with two other CDM
problems.  The solution of the long--standing angular--momentum problem in
galaxy formation might be solved if ``subclumps'' in the protogalaxy's halo
were eliminated (CDM models produce too small disks, thought to be attributable
to angular momentum transfer from the infalling gas to the DM subclumps).
Although WDM does suppress halo substructure, the required WDM mass is
estimated to be $0.5-0.8$ keV \cite{larsen}.  Furthermore, a simulation that
artificially turns off angular momentum transfer between gas and DM halo
subclumps reveals that the specific angular momentum distribution still
contradicts observations \cite{kravtsov}. Finally, the absence of dwarf
galaxies in voids, a difficulty of CDM \cite{peeblesvoid}, is found to be eased
in a large--scale simulation of an $m_X=1.5$ keV WDM model \cite{Bode}, but a
smaller $m_X$ appears necessary to match the observations in detail.

\begin{figure}[b!] % fig 1
\centerline{\epsfig{file=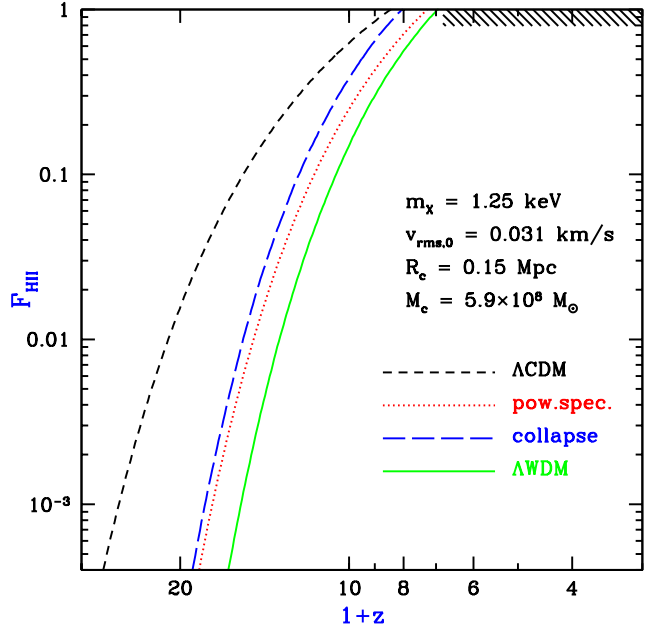,height=3.5in,width=3.5in}}
\vspace{10pt}
\caption{The filling factor of ionized hydrogen, $F_{\rm H\,II}$, as 
a function of redshift $z$ in our standard model ($C=10,
\epsilon_*=0.01$). The uppermost curve corresponds to
$\Lambda$CDM, and the lowest curve to a WDM particle mass of
$m_X=1.25$ keV. The middle pair of curves shows separate contributions
to the delay in reionization from the suppression of the power
spectrum (dotted curve) and the ``effective pressure'' of WDM (long
dashed curve). 
%The shaded region indicates the requirement $z_{\rm
%reion}>5.8$.
}
\label{fig:reion}
\end{figure}

\begin{figure}[b!] % fig 2
\centerline{\epsfig{file=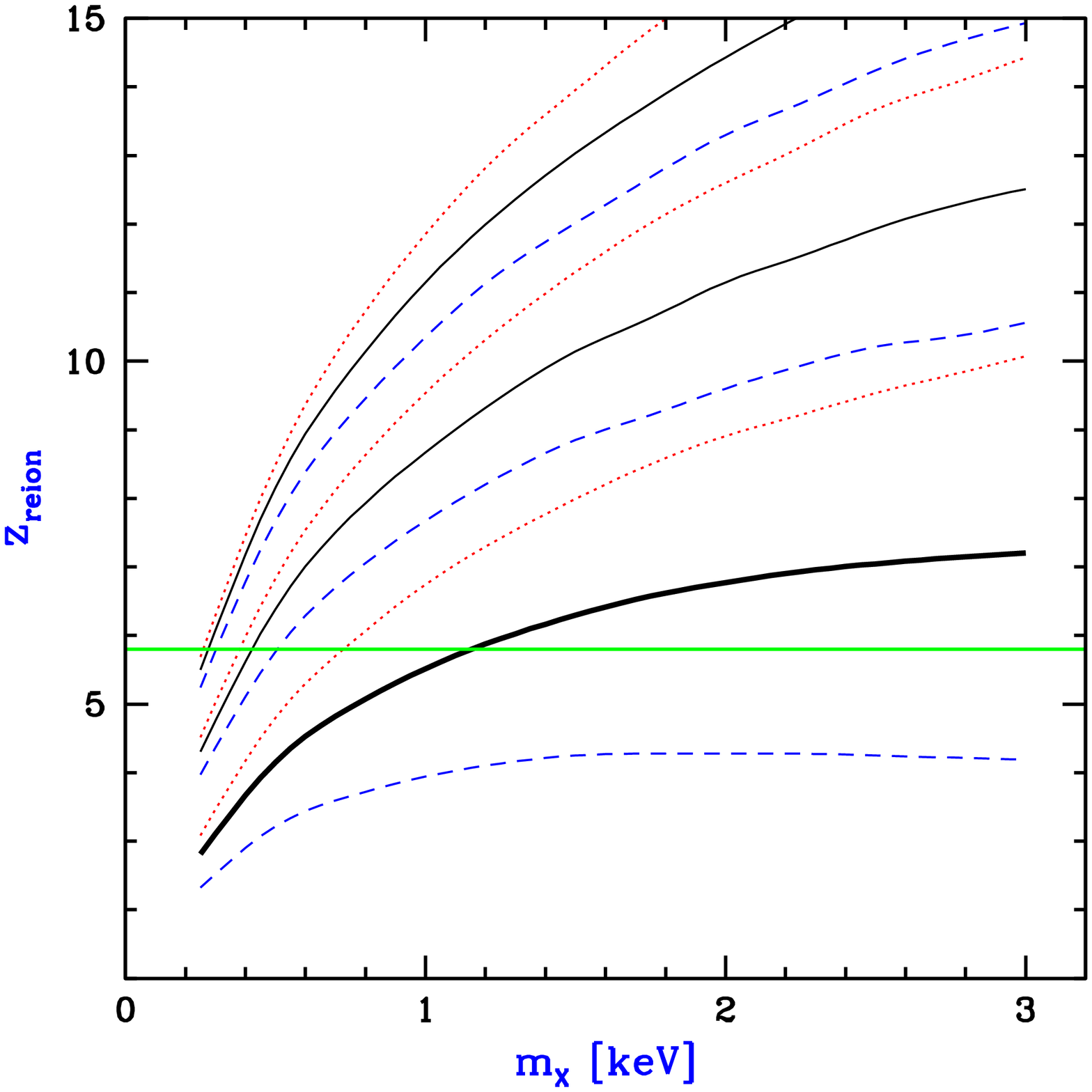,height=3.5in,width=3.5in}}
\vspace{10pt}
\caption{Reionization redshift as a function of WDM particle mass. 
Our standard model for the star-formation efficiency, escape fraction,
and stellar IMF, is shown by the thick solid line.  The three solid
curves show models with the same clumping factor ($C=10$) but
different ionization efficiencies ($\epsilon_*=0.01,0.1,1$, bottom to
top).  The three dotted curves correspond to the same models except
with $C=1$; the three dashed curves correspond to the same models 
with $C=30$.  In our standard model, the requirement $z_{\rm
reion}>5.8$ yields the constraint $m_X > 1.25$~keV.}
\label{fig:zreion}
\end{figure}

\section*{Constraints from Reionization}  

%In attempting to solve the small--scale problems of CDM, it is
%important not to introduce new problems.  
As emphasized in section \S~1, WDM models reduce the abundance of
high--redshift halos, which could lead to conflicts with observations.
Note that in general, the differences between WDM and CDM models are
amplified at high redshifts. In a study based on a modified version of
the extended Press--Schechter theory, we have quantified constraint on
the velocities and masses of WDM particles from reionization, and from
the presence of a supermassive BH in the bright $z=5.8$ quasar SDSS
1044-1215. The main features of our models are as follows:
\begin{enumerate}
\item The effect of free streaming on the power spectrum 
was included using a fit to the numerical transfer function, obtained
from a Boltzmann code.
\item The effect of WDM velocities on the growth of perturbations 
was included using spherically symmetric, one--dimensional
hydrodynamical simulations.
\item The WDM halo mass function was computed using Monte Carlo--generated
halo ``merger trees''. A moving barrier ($\delta_c\geq 1.69$) was
adopted to generalize (an improved) Press--Schechter theory to the
case of mass--dependent collapse times.  Our semi--analytic
halo mass function was demonstrated to agree with numerical
simulations at low redshift over the halo masses of interest
\cite{Bode}.
\item We assume that a fraction $f_*$ of baryons turns into stars in halos with
virial temperatures $T_{\rm vir}\geq10^4$K (necessary for efficient cooling
\cite{har00}). We assume a Scalo (1998) stellar IMF (producing $\approx 4000$
ionizing photons per baryon), and that a fraction $f_{\rm esc}$ of the ionizing
photons escape into the IGM.  We parameterize our models by the product
$\epsilon_* \equiv f_* f_{\rm esc} ( =0.01$ in our standard model$)$,
consistent with the ionizing background inferred from the proximity effect
at redshift $z\approx 3$. \cite{bdo}
\item The mean clumping of the ionized gas in the IGM is parameterized
by a single constant $C\equiv \langle n_{\rm H}^2 \rangle / \bar
n_{\rm H}^2 ( =10$ in our standard model$)$.
\end{enumerate}

Our main results on the constraints from reionization are shown in
Figures \ref{fig:reion} and \ref{fig:zreion}. Figure \ref{fig:reion}
shows the filling factor of ionized hydrogen in our standard model,
with $m_X=1.25$ keV, $\epsilon_*=0.01$, and $C=10$.  In this model,
reionization occurs at $z=5.8$. The contributions to the decrease of
the filling factor from the suppression of the power spectrum and the
effective pressure of WDM are comparable (dotted and long--dashed
curves, respectively). In Figure \ref{fig:zreion}, we demonstrate how
the reionization redshift depends on $m_X$, and on our two model
parameters, $\epsilon_*$ and $C$.

\section*{Conclusions}  

If high-redshift galaxies produce ionizing photons with an efficiency similar
to their $z=3$ counterparts ($\epsilon_*\sim0.01$), reionization by redshift
$z=5.8$ places a limit of $m_X\ga 1.25~$ keV ($\vzrms \la 0.03$ km/s) on the
mass of the WDM particles. This limit is somewhat stronger than the limit
inferred from the statistics of the Ly$\alpha$ forest (which yields $m_X\ga
0.75~$keV; \cite{vijay00}), although our limit may weaken to $m_X\ga0.75~$keV
($\vzrms = 0.060$ km/s) given the uncertainty in current measurements of the
stellar contribution to the ionizing intensity at $z=3$ \cite{bdo} (which we
can use to normalize our models). We also find that the existence of a $\approx
4\times 10^9~{\rm M_\odot}$ supermassive black hole at $z=5.8$, powering the
quasar SDSS 1044-1215 (assuming it is unlensed and radiating at or below the
Eddington limit), yields the somewhat weaker, but independent constraint
$m_X\ga 0.5$ keV (or $\vzrms \la 0.10$ km/s), if this BH acquired most of its
mass accreting at the Eddington rate.  Finally, we also find that WDM models
with $m_X\la1~$keV ($\vzrms \ga 0.04$ km/s) produce a low-luminosity cutoff in
the high-$z$ galaxy luminosity function which is detectable with the {\it Next
Generation Space Telescope}. Such an observation would directly break the
degeneracy in the reionization redshift between low ionizing-photon production
efficiency and small WDM particle mass.  The constraints derived here will
tighten considerably as observations probe still higher redshifts, offering
increasingly stringent tests of models with diminished small--scale power,
exemplified by WDM.

\vspace{\baselineskip} ZH acknowledges support from a Hubble Fellowship, and RB
acknowledges support from CITA and from Institute Funds (IAS, Princeton).

\end{document}